\documentclass[lettersize,journal]{IEEEtran}
\usepackage{amsmath,amsfonts}
\usepackage{algorithmic}
\usepackage{algorithm}
\usepackage{array}
\usepackage[caption=false,font=normalsize,labelfont=sf,textfont=sf]{subfig}
\usepackage{textcomp}
\usepackage{stfloats}
\usepackage{url}
\usepackage{ulem}
\usepackage{verbatim}
\usepackage{graphicx}
\usepackage{subfig}
\usepackage{makecell}
\usepackage{cite}
\begin{document}

\title{Towards Integrated Fine-tuning and Inference when Generative AI meets Edge Intelligence}
%
\author{Ning~Chen,~\IEEEmembership{Student Member,~IEEE, } 
       Zhipeng~Cheng,~\IEEEmembership{Member,~IEEE, }
       Xuwei~Fan,~\IEEEmembership{Student Member,~IEEE, }
       Xiaoyu~Xia,~\IEEEmembership{Member,~IEEE, }
       Lianfen~Huang~\IEEEmembership{}\\
\thanks{Corresponding author:  Zhipeng Cheng (e-mail: chengzp\_x@163.com)}
\thanks{Ning Chen, Xuwei Fan, and Lianfen Huang are with the School of Informatics, Xiamen University, 361005 Xiamen, China (email: ningchen@stu.xmu.edu.cn; xwfan@stu.xmu.edu.cn; lfhuang@xmu.edu.cn).}
\thanks{Zhipeng Cheng  is with the School of Future Science and Engineering, Soochow University, 215006 Suzhou, China (email: chengzp\_x@163.com).}
\thanks{Xiaoyu Xia is with the School of Computing Technologies, RMIT University, Melbourne, VIC 3000, Australia (e-mail: xiaoyushaw@gmail.com).}

}



\maketitle

\begin{abstract}
The high-performance generative artificial intelligence (GAI) represents the latest evolution of computational intelligence, while the blessing of future 6G networks also makes edge intelligence (EI) full of development potential. The inevitable encounter between GAI and EI can unleash new opportunities, where GAI's pre-training based on massive computing resources and large-scale unlabeled corpora can provide strong foundational knowledge for EI, while EI can harness fragmented computing resources to aggregate personalized knowledge for GAI. However, the natural contradictory features pose significant challenges to direct knowledge sharing. To address this, in this paper, we propose the \uline{GAI}-oriented \uline{s}ynthetical \uline{net}work (GaisNet), a collaborative cloud-edge-end intelligence framework that buffers contradiction leveraging data-free knowledge relay, where the bidirectional knowledge flow enables GAI’s virtuous-cycle model fine-tuning and task inference, achieving mutualism between GAI and EI with seamless fusion and collaborative evolution. Experimental results demonstrate the effectiveness of the proposed mechanisms. Finally, we discuss the future challenges and directions in the interplay between GAI and EI.
\end{abstract}


\section{Introduction}
\subsection{Motivation}
\IEEEPARstart{G}{enerative} artificial intelligence (GAI) as the key enabling technology of artificial intelligence-generated content (AIGC) has been rapidly developed and applied in various tasks, e.g., natural language processing (NLP), computer vision (CV), code generation, and mathematical reasoning  \cite{ref1, ref2, ref18}. Table \ref{table_1} shows some well-known GAI foundation models (FMs) \cite{ref15}, which are characterized by large parameter scales, common knowledge learning based on the pre-training on the massive unlabeled corpus, and domain-across robustness of few-shot fine-tuning \cite{ref7}. However, studies have shown that existing centralized high-quality language data is expected to be exhausted by 2026, and even low-quality language and image data will be exhausted by 2030 to 2050, and GAI faces the challenge of limited availability of high-quality public data \cite{ref7}. Therefore, to sum up, although it has a strong performance brought by foundation knowledge, the shortage of high-quality legal data and the unaffordable computing resources brought by large-scale parameters lead to the centralized development under the monopoly of the Internet giants, which hinders the diversification and democratization of GAI.

\begin{table*}[t]
\center
\footnotesize
\caption{Some Representative GAI Foundation Models.}
\label{table_1}
\renewcommand\arraystretch{1.5}
\begin{tabular}{|c|c|c|c|c|c|}
\hline
& \textbf{Publisher}  & \textbf{Architecture} & \textbf{Size} & \textbf{Data} & \textbf{Modality} \\ \hline
\textbf{GPT-3} & OpenAI & Decoder & 175 B & 300 B (Tokens) & Text \\ \hline
\textbf{GPT-4} & OpenAI & Decoder & 170 T & 10 T (Tokens) & Text, Image \\ \hline
\textbf{LLaMA} & Meta & Decoder & 7 B-65 B & 1.4 T (Tokens) & Text \\ \hline
\textbf{BERT} & Google & Encoder & 110 M, 335 M & 250 B (Tokens) & Text \\ \hline
\textbf{DALL-E} & OpenAI & VAE, Decoder & 12 B & 2.5 B (Pairs) & Text, Image \\ \hline
\textbf{DALL-E 2} & OpenAI & CLIP, Diffusion & 3.5 B & 6.5 B (Pairs) & Text, Image \\ \hline
\end{tabular}
\end{table*}

On the other hand, different from centralized GAI with large-scale parameters, edge intelligence (EI) is more inclined to deploy flexible lightweight models around users, making computational intelligence closer to distributed terminal data \cite{ref19,ref22,ref23}. According to a recent analysis report, there will be 30.9 billion Internet of Things (IoT) devices connected in 2025, and the data scale is expected to reach nearly 79.4 Zettabytes (ZB) \cite{ref21}. Meanwhile, due to the continuous upgrading of chip integration technologies such as central processing unit (CPU) and graphics processing unit (GPU) on mobile terminals, the power consumption cost is continuously reduced, which endows terminal equipment with certain computing capabilities for the lightweight AI model training and inference \cite{ref19}. Therefore, benefiting from the growth of edge data and the enhancement of computing power of terminal devices, in the coming B5G and 6G era, we will witness the transformation of the network paradigm from the Internet of everything to the intelligence of everything, in which native AI will sink from distant cloud servers to the edge of the network \cite{ref11}. However, even if EI is closer to terminals, the limited model scale leads to the lack of prior knowledge, which makes the effect of edge training and inference unsatisfactory.

Based on the above analysis, GAI and EI can be combined to achieve complementary advantages, and the interplay between them will derive new potential growth energy. Harnessing personalized knowledge by fragmented computing resources, EI can alleviate the dilemma of public data shortage of GAI and bridge the distance between computational intelligence and end users, while GAI can endow EI with pre-trained generalized foundation knowledge, which can be set as the robust baseline for EI to accelerate learning convergence and improve inference performance \cite{ref7,ref11}. However, as shown in TABLE \ref{table_2}, due to the contradictory characteristics of parameter size, application domain, network architecture, data size, and resource provision, the knowledge transfer between GAI and EI is greatly hindered, which can be summarized as follows.

$\bullet$ \textbf{Interruption of uplink knowledge between EI and GAI caused by the blocked data pipeline.} Firstly, limited by the insufficient communication resources of terminals, the exponentially increasing local data cannot be uploaded to the cloud, which makes the knowledge pipeline from EI to GAI blocked. In addition, the vanilla GAI models are usually deployed in the central cloud of large enterprises or research institutions, such as Amazon Cloud, and use the massive collected data to perform the pre-training and fine-tuning. However, to avoid privacy leakage, 6G terminal devices are usually unwilling to share local data with the GAI model in the cloud, which also brings great obstacles for the GAI to utilize personalized knowledge of 6G terminal data \cite{ref10}.

$\bullet$ \textbf{Interruption of downlink knowledge between GAI and EI caused by the blocked model pipeline.} GAI relies on large-scale models with billions of parameters pre-trained on large-scale basic datasets, which requires huge computing and time resources \cite{ref11,ref12,ref18}. For example, training a GPT-3 model with more than 175 billion parameters requires 1000 GPUs running more than 4 months \cite{ref11}. To train the Stable Diffusion model, Stability AI maintains over 4000 NVIDIA A100 GPU clusters, spending over $\$$50 million in operating costs \cite{ref12,ref18}. Similarly, in the task inference scenario, the memory requirement increases dramatically with the number of parameters. For example, Falcon-40B requires about 86 GB of GPU memory for inference \cite{ref15}. Whether for training or inference, such a large resource cost is burdensome for the resource-constrained 6G terminals, which leads to the blockage of the transfer pipeline of foundation knowledge from GAI to EI.

\begin{table*}[t]
\center
\footnotesize
\caption{The Crucial Contradictory Characteristics between GAI and EI.}
\label{table_2}
\renewcommand\arraystretch{1.5}
\begin{tabular}{|c|c|c|c|c|c|}
\hline
& \textbf{Size} & \textbf{Domain} & \textbf{Architecture} & \textbf{Data} & \textbf{Resource} \\ \hline
\textbf{GAI} & Heavyweight & Generalized & Centralized & Massive & Sufficient \\ \hline
\textbf{EI} & Lightweight & Personalized & Distributed & Slight & Limited \\ \hline
\end{tabular}
\end{table*}

Facing the above challenges, hybrid federated splitting learning (HFSL), which retains the data of 6G terminal devices locally and performs GAI model segmentation, can provide a feasible solution, where federated learning (FL) is a collaborative learning paradigm that aggregates personalized knowledge of terminals for GAI without data sharing \cite{ref7,ref13,ref19}, while split learning (SL) can split GAI model and deploy resource-adapted lightweight sub-modules on terminals to perform collaborative training and inference \cite{ref12,ref13}. However, due to the generalization characteristics of GAI’s FMs, it is crucial to perform domain-specific model fine-tuning before applying it to perform task inference, and the original FMs cannot provide GAI task inference service with optimal performance. Meanwhile, the one-way development from development to application of GAI is a poor and inefficient pattern, and the growth of the terminal cannot be fed back to the initial model. Therefore, it is necessary to develop a sustainable-evolution GAI-oriented network architecture that can realize both model fine-tuning and task inference.

\subsection{Novelty and Contributions}

To solve the above problems, first, fully respecting the resource imbalance, we propose the \uline{GAI}-oriented \uline{s}ynthetical \uline{net}work (GaisNet), a collaborative cloud-edge-end intelligence framework tailored for GAI’s model fine-tuning and task inference. To the best of our knowledge, this paper is among the first to study the interplay between GAI and EI from the perspective of integrated fine-tuning and inference. Major contributions are summarized as follows.

$\bullet$ GaisNet, a collaborative cloud-edge-end intelligence framework, is proposed, where the domain-specific edge model acts as the data-free knowledge relay, which unlocks the bidirectional knowledge flow between GAI and EI, realizing the sustainable-evolution model fine-tuning and task inference.

$\bullet$ We enumerate and analyze the major issues that will be encountered in the operation of GaisNet with integrated fine-tuning and inference.

$\bullet$ Experimental results explore the influencing factors of GaisNet and demonstrate the effectiveness of the proposed mechanisms.

$\bullet$ Future challenges and directions in the interplay between GAI and EI are discussed.

The rest of this article is organized as follows. In Section II, the preliminaries are introduced, and then we propose the GaisNet framework in Section III, and then the major issues for GaisNet is discussed in Section IV. The simulation results are discussed and analyzed in Section V. Future opportunities and directions regarding the interplay of GAI and EI are discussed in Section VI, before drawing the conclusion in Section VII.

\section{Preliminaries}

\subsection{Generative AI and Foundation Model}

Benefiting from the breakthroughs in various underlying backbones, e.g., generative adversarial networks (GAN), variational auto-encoders (VAE), diffusion model (DM), and Transformer[11], technology giants around the world have released a variety of powerful FMs, that is, the universal GAI models obtained from pre-training based on massive data and computing resources, e.g., OpenAI's GPT series and DALL-E series, Meta's LLaMA, Google's BERT and PaLM \cite{ref15,ref1,ref7}.

The lifecycle of the GAI model is composed of \textit{pre-training}, \textit{fine-tuning}, and \textit{inference}. First, the domain-across pre-training endows the FMs with the generalized foundation knowledge. Then, the domain-specific model fine-tuning transforms the output of pre-training towards a narrower range of professional foundation knowledge following human intentions. Last, task inference is the application and ultimate goal of the pre-trained and fine-tuned GAI model \cite{ref11,ref8}. Below we discuss what each process does and what challenges it faces.

1) \textbf{Pre-training} is the main optimization method of GAI models. Pre-training on large-scale unlabeled datasets with self-supervised representation endows the GAI model with domain-across generalized foundation knowledge. However, the pre-training with billions of parameters requires a great amount of computing and time resources \cite{ref12}, which makes it difficult for most resource-constrained terminal devices to directly participate in the pre-training process, resulting in the waste of personalized local data of edge devices \cite{ref11}.

2) \textbf{Fine-tuning} is the re-optimization of the GAI model after pre-training. The fine-tuning for different vertical domains makes the original GAI model further increase the domain-specific knowledge and further improve its adaptability to the downstream data. Better coping with downstream tasks in different domains [8]. The fine-tuning methods of the GAI model can be divided into full fine-tuning which updates all backbone parameters and parameter-efficient fine-tuning which only updates the lightweight tunable modules. Thus, parameter-efficient fine-tuning can reduce the number of trainable parameters by freezing the model backbone, thereby alleviating the pressure on storage memory and computing resources. The state-of-the-art parameter-efficient fine-tuning methods include preﬁx tuning, adapter tuning, low-rank adaptation (LoRA), etc \cite{ref5,ref4,ref6,ref9}. The vanilla centralized fine-tuning relies on continuous data collection from users, which violates users' personal data privacy \cite{ref11}.

3) \textbf{Inference} is more biased to the application of the optimized GAI model. Terminals input unlabeled data (e.g., prompt instruction and images, etc.) into the pre-trained and fine-tuned GAI model, and the GAI model outputs the content following the user intention, such as reply text and drawn images, so as to meet users' service requirements \cite{ref19}. The inference is affected by the domain relevance of intent and model and the uncertainty of the network environment \cite{ref11}.

Thus, \textit{model fine-tuning} is necessary to align the pre-trained FM to human intention and produce personalized output, and \textit{task inference} is the fundamental way to test the performance of GAI models. Although GAI’s FM is high-performance, its large parameter scales lead to the closed-form optimization and application led by the head enterprise, which makes GAI far away from the massive resource-constrained terminal clients, so that the widely distributed fragmented computing resources and personalized local data cannot be fully utilized. On the contrary, EI can sink AI to the user side, making distributed clients more involved in AI-related work.

\subsection{Edge Intelligence and Distributed learning}

Cloud computing can provide sufficient computing power for various terminal applications, and play an important role in the training and inference of AI models \cite{ref21}. However, with the increasing terminal data, AI based on full cloud computing faces significant challenges in meeting latency limitations in real-time applications such as autonomous driving and telemedicine, while the risk of privacy leakage also brings great concerns. The traditional cloud-based data-centralized model training and inference are unable to meet the surging data traffic demand, ubiquitous computing demand, strict delay limitation, and personalization requirements of artificial intelligence of things (AIoT) applications \cite{ref13}. For the above problems, EI can provide low-latency AI services for mobile terminals by subsiding the deployment, training, and inference of computational intelligence to the user side and utilizing the computing resources and personalized data scattered at the edge of the network, thus solving the “last kilometer” problem of AI \cite{ref19}. Meanwhile, due to the famous Moore's law, the computing capacity of terminal clients is becoming more and more powerful, which can support the localized running of machine learning (ML) tasks, creating conditions for the development of distributed learning \cite{ref19}.

From the perspective of data processing, EI includes centralized EI with data aggregation and distributed EI without data aggregation. Among them, the major representative frameworks of distributed EI are FL, SL, and HFSL, which can realize the learning of data knowledge and finish the inference tasks under the premise of privacy protection \cite{ref7,ref14}. FL is a collaborative learning paradigm, in which local data transmission is replaced by model aggregation and distribution, which can indirectly utilize the knowledge of distributed data under the premise of protecting user privacy \cite{ref7,ref13,ref19}. SL can further decompose the complex AI model and only deploy the lightweight part peer to the terminal client’s resource \cite{ref12,ref13}. Although EI builds the association between AI and local data of terminals and realizes more accessible training and inference processes closer to users, it is limited by the shortage of multi-domain physical resources such as computing resources required for model training and inference and communication resources required for model parameter transmission. As a result, it is difficult for EI to obtain high-quality model training and task inference. In this case, collaborative cloud-edge-end intelligence, which can organically connect the high-performance models in the cloud and the terminals’ lightweight models, can provide us with ideas to solve the above problem.

\subsection{Collaborative Cloud-edge-end Intelligence}
Driven by the vigorous development of AI and distributed computing paradigms, it is urgent to combine distributed AI with hierarchical computing networks \cite{ref21}. In addition, with the appearance of heterogeneous networking architectures such as 5G networks with macro base stations and small cells coexisting, the communication pipeline between terminal clients, edge access points (e.g., small cells, roadside units (RSUs), etc.) and macro base stations is opened up, and the physical feasibility of collaborative cloud-edge-end intelligence is constructed. Following the requirements of AI services, the computing and communication resources of the hierarchical heterogeneous cloud-edge-end network can be reasonably arranged and utilized, and the rich computing resources of cloud servers and the low access latency of edge servers were utilized to support computation-intensive training and delay-sensitive inference \cite{ref21}. Fully respecting the resource imbalance on different components, collaborative cloud-edge-end intelligence can realize the suitable work allocation. First, the \textit{cloud layer} has abundant computing power and rich storage resources, which can achieve high-performance model training and task inference. Then, the \textit{edge layer} is a bridge connecting the cloud and the terminals, which can realize low-latency model training and task inference. Last, the \textit{end layer} is a data source with a certain amount of computing and storage capacity that can constantly generate data, which can realize lightweight model training and task inference.

\section{GaisNet with Integrated Fine-tuning and Inference}

\subsection{Parameter-efficient Fine-tuning and Inference}

Parameter-efficient fine-tuning has been widely studied, which mainly focuses on efficient domain-specific model re-training from the computing perspective. First, we discuss parameter-efficient fine-tuning by prompt tuning. Then, with a similar concept, we propose parameter-efficient inference from the communication perspective, which only transmits small-scale tunable modules while keeping the backbone with large-scale parameters frozen. As a result, the communication overhead of parameter transmission is reduced.

\textit{1) Computing Perspective: Parameter-efficient Fine-tuning}

\begin{figure}[t]
\centering
\includegraphics[width=3.2in]{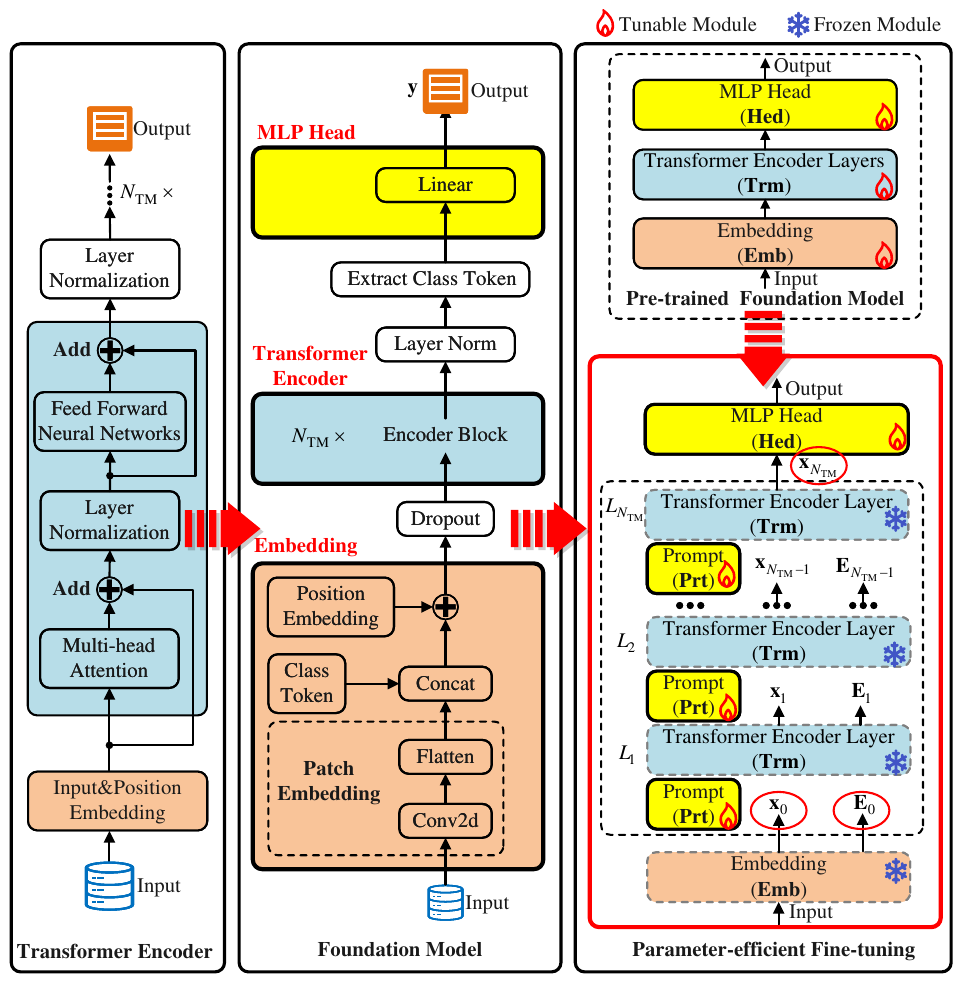}
\caption{Transformer-based foundation model and parameter-efficient fine-tuning.}
\label{fig_1}
\end{figure}

The success of GAI relies on the development and innovation of backbone architectures, the most famous of which is the Transformer proposed by Google in 2017 \cite{ref2,ref1,ref9}. Transformer is the backbone of many state-of-the-art FMs, e.g., GPT-3, DALL-E 2, Codex, and Gopher. The basic structure of the Transformer is shown in Fig. \ref{fig_1}. For simplicity and without loss of generality, in this paper, we discuss the GAI model based on Transformer as the underlying model. As shown in Fig. \ref{fig_1}, pre-trained FMs based on Transformer encoder architecture, e.g., BERT, ViT, etc., can be divided into the embedding layer (Emb) for feature extraction, the multiple Transformer layers (Trm) for self-attention-based learning, and the multilayer perceptron (MLP) head layer (Hed) for output judgment \cite{ref3}.

A key step in applying pre-trained FM models to various downstream tasks is fine-tuning \cite{ref4,ref5}. As a representative parameter-efficient fine-tuning method, \textit{prefix tuning} based on prompt learning adds some learnable prompt modules in front of each Transformer layer. Compared with full fine-tuning, using downstream data to train the prompt modules and freezing the entire pre-trained Transformer backbone can achieve domain performance enhancement, just like the "Four ounces can move a thousand pounds". For example, visual-prompt tuning (VPT) proposes an efficient parameter-efficient fine-tuning method for vision transformer (ViT) based on prompt learning, which achieves desirable performance in vision tasks. Only a small number (less than 1$\%$ of the model parameters) of trainable parameters used to control the prompt modules are introduced in the input space, which is learned together with the linear MLP head during fine-tuning \cite{ref3,ref4,ref9}.

As shown in Fig. \ref{fig_1}, consider an FM with ${{N}_{\text{TM}}}$ Transformer encoder blocks \cite{ref3}, and the prompts are introduced into the input space of each Transformer layer \cite{ref4}. Define ${{\mathbf{E}}_{0}}$ and ${{\mathbf{x}}_{0}}$ are the initial embeddings of the input data patches and learnable classification tokens $\left[ \text{CLS} \right]$, respectively. Thus, the output tokens of the $i\text{-th}$ Transformer layer can be expressed as

\begin{equation}\label{e1}
\left[ {{\mathbf{x}}_{i}},{{\mathbf{E}}_{i}} \right]={{\mathcal{L}}_{i}}\left( \left[ {{\mathbf{x}}_{i-1}},{{\mathbf{E}}_{i-1}} \right] \right),\text{ }1\le i\le {{N}_{\text{TM}}}
\end{equation}

After the self-attention learning of the ${{N}_{\text{TM}}}$ Transformer layers, the learnable classification token from the output of the ${{N}_{\text{TM}}}\text{-th}$ Transformer layer is fed into the MLP head to output the final result, i.e.,

\begin{equation}\label{e2}
\mathbf{y}=\mathcal{H}\left( {{\mathbf{x}}_{{{N}_{\text{TM}}}}} \right)
\end{equation}

Finally, after the loss calculation and gradient feedback, the fine-tuning process of FM on the data of a specific field further improves FM’s performance in the relevant applications.

\textit{2) Communication Perspective: Parameter-efficient Task Inference}

To improve the model utilization and task inference efficiency of GAI, for the service requirements of similar inference tasks, transfer learning based on parameter sharing can realize the shared application of the fine-tuned models. We define parameter-full inference as fully sharing all parameters and the parameter-efficient inference in which only the lightweight module parameters are shared, of course, which is defined from the communication perspective, like parameter-efficient fine-tuning from the computing perspective.

\begin{figure}[t]
\centering
\includegraphics[width=2.8in]{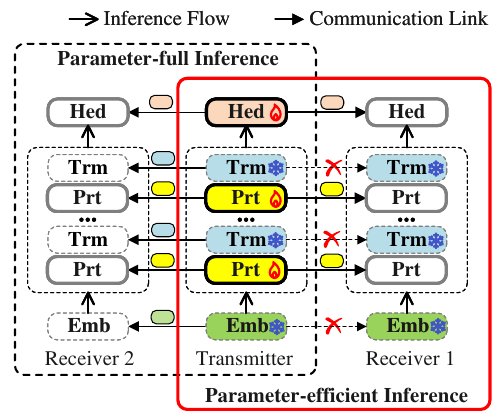}
\caption{Parameter-efficient inference from the communication perspective.}
\label{fig_2}
\end{figure}

Benefiting from parameter-efficient model fine-tuning, we enable parameter-efficient task inference in scenarios involving model distribution and reconfiguration. The model distribution only involves the updated lightweight modules, while keeping the backbone frozen, which reduces the communication overhead caused by transferring large-scale model parameters or smashed data (e.g., intermediate activations, back-propagated gradients) \cite{ref12,ref13}. As shown in Fig. \ref{fig_2}, the transmitter shares the fine-tuned model with both receivers. For the parameter-full inference scenario, all the model parameters, including the parameters of the large-scale Transformer module, need to be transmitted. For the parameter-efficient inference, only the parameters of the MLP head and prompt modules that can be re-trained during the fine-tuning process are transmitted to the receiver, which greatly reduces the high communication overhead and inference delay caused by parameter sharing.

\subsection{GAI-oriented Synthetical Network (GaisNet)}

\begin{figure*}[t]
\centering
\includegraphics[width=5.5in]{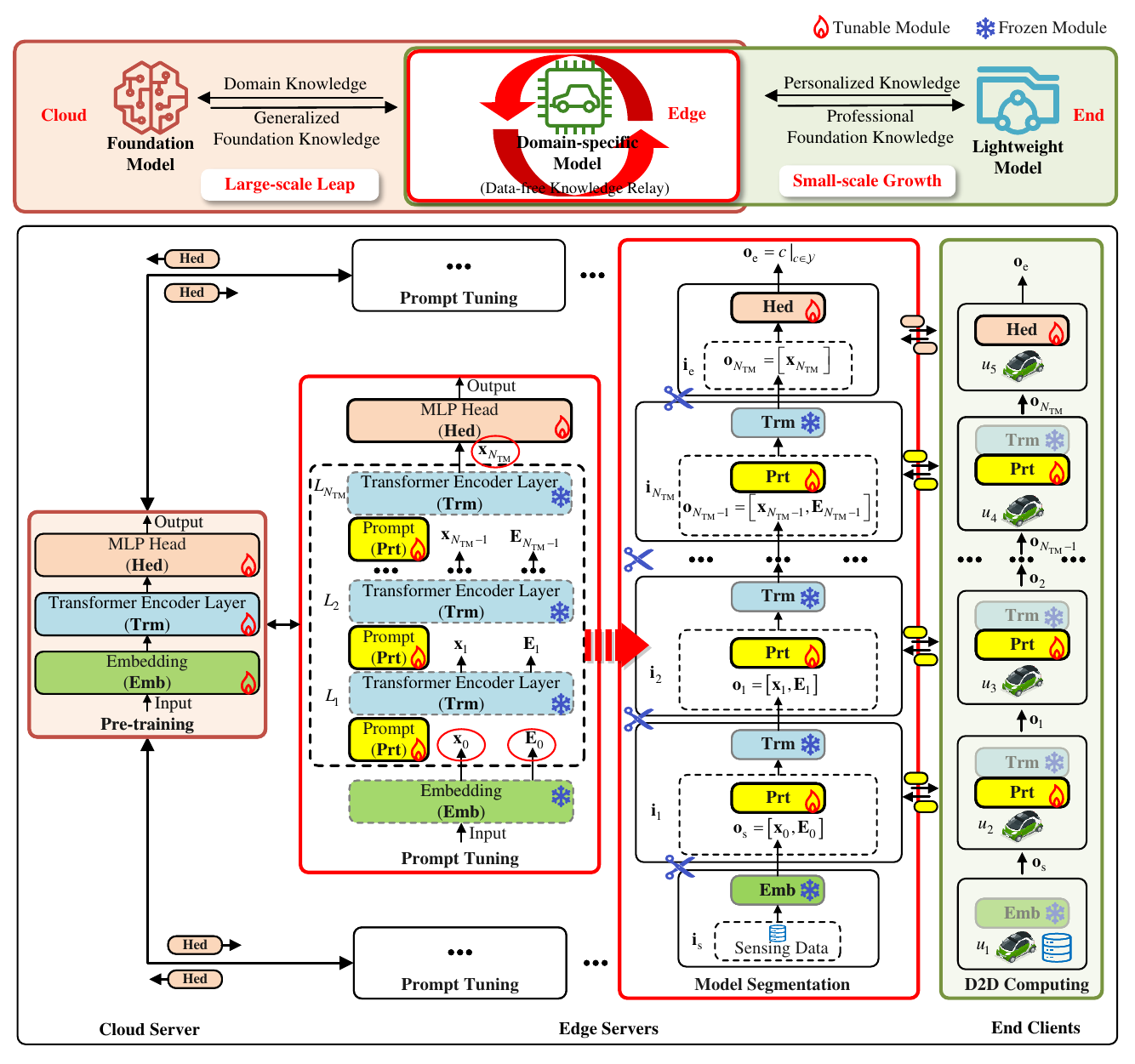}
\caption{The architecture of the proposed GaisNet.}
\label{fig_3}
\end{figure*}

Fig. \ref{fig_3} shows the architecture of the proposed GaisNet, a collaborative cloud-edge-end intelligence framework for GAI model fine-tuning and task inference, which consists of foundation models deployed on cloud servers, domain-specific models deployed on edge servers, and lightweight models deployed in 6G terminal clients. GaisNet is built upon the collaborative cloud-edge-end intelligence framework and the D2D communication framework. After segmentation, the tunable modules of the edge model are transferred to the client cluster, where the embedding layer is deployed at the start point and its output is ${{\mathbf{o}}_{\text{s}}}=\left[ {{\mathbf{x}}_{0}},{{\mathbf{E}}_{0}} \right]$, while the MLP head layer is deployed at the end point, and the output (i.e., the final classification result) is defined as ${{\mathbf{o}}_{\text{e}}}={{\left. c \right|}_{c\in \mathcal{Y}}}$, where the alphabet $\mathcal{Y}=\left\{ {{y}_{1}},{{y}_{2}},\cdots ,{{y}_{|\mathcal{Y}|}} \right\}$ represents the label space, such as the set of classes in object classification or recognition tasks. In addition, the middle ${{N}_{\text{TM}}}$ Transformer layers are deployed on the remaining clients of the cluster.

Pre-training based on the massive computing resources of cloud servers and large-scale unlabeled corpora endows FMs with domain-across generalized foundation knowledge, which is further converted into domain-specific professional foundation knowledge by fine-tuning for better adaptation to different downstream tasks. It is worth noting that in this paper we focus on parameter-efficient model fine-tuning and task inference, i.e., only re-training and transmitting the tunable part of the pre-trained FMs, and it is assumed that the frozen backbones (e.g., the large-scale Transformer modules and the Embedding module for feature extraction) are always synchronized independently in the cloud server, the edge servers and the end clients, which are not under the optimization consideration in this paper.

In particular, the edge server associates the cloud server and end clients in the role of data-free knowledge relay, realizing bidirectional knowledge flow for model fine-tuning and task inference, where the cloud-edge-end collaborative GaisNet is further divided into cloud-edge subnetworks with domain-across large-scale knowledge flow and edge-end subnetworks with domain-specific small-scale knowledge flow.

$\bullet$  \textbf{Cloud-edge subnetworks with domain-across large-scale knowledge flow.} The FL framework is adopted to perform domain-across distributed learning between a single cloud service and multiple edge servers, where the cloud server uses the large-scale unlabeled corpora and the full-model pre-training with massive computing resources to give the GAI FMs generalized foundation knowledge, which will be transferred to the domain-specific models of the edge server in the process of cloud model delivering. Meanwhile, the edge server aggregates the personalized knowledge learned from distributed terminal clients to obtain the domain knowledge, which will be transferred to the FMs of the cloud server in the process of edge model parameter uploading, further enhancing the FMs’s performance of cross-domain generalization. Therefore, the bidirectional knowledge flow of the cloud-edge subnet can form a benign closed loop between FMs and domain-specific models.

$\bullet$  \textbf{Edge-end subnetworks with domain-specific small-scale knowledge flow.} Due to the data sparseness of edge servers and the clients’ privacy considerations and resource constraints, to make full use of the data and computing resources of distributed clients, the clients’ lightweight models are used to jointly perform HFSL-based model fine-tuning and SL-based task inference. The edge server associates domain-related clients with cell-free networking and provides them with the domain-specific foundation knowledge obtained from the cloud server and fine-tuned toward the domain data. Meanwhile, the personalized knowledge in clients’ local data is collected to obtain the new domain knowledge. Thus, the bidirectional knowledge flow of the edge-end subnet can form a benign closed loop of mutualism between the edge domain-specific model and the terminals’ lightweight model.

With the collaboration among the FMs deployed in the cloud server, the domain-specific models of the edge servers, and the local lightweight models of the end clients, it can break the blocked knowledge flow caused by the data blocking from EI to GAI and the model blocking from GAI to EI, where GAI provides domain-across foundation knowledge for EI through the high-performance FMs, while EI provides the domain-specific personalized knowledge to GAI through the distributed edge models. As a data-free bidirectional knowledge relay, the edge server can not only transfer the large-scale enhanced popular knowledge obtained from the large dataset of the cloud server, where not only the domain-specific task inference services can be offered for end clients but also collects the personalized knowledge from the terminal clients' local data in a privacy-preserving way. Therefore, GaisNet can achieve a benign closed-loop GAI model evolution, which can further achieve sustainable and efficient model fine-tuning and task inference.

Then, since edge-end subnets are more complex and representative, we focus on its two subprocesses, i.e., \textit{HFSL-based model fine-tuning} and \textit{SL-based task inference}.

\subsection{HFSL-based Model Fine-tuning}

\begin{figure}[t]
\centering
\includegraphics[width=3.0in]{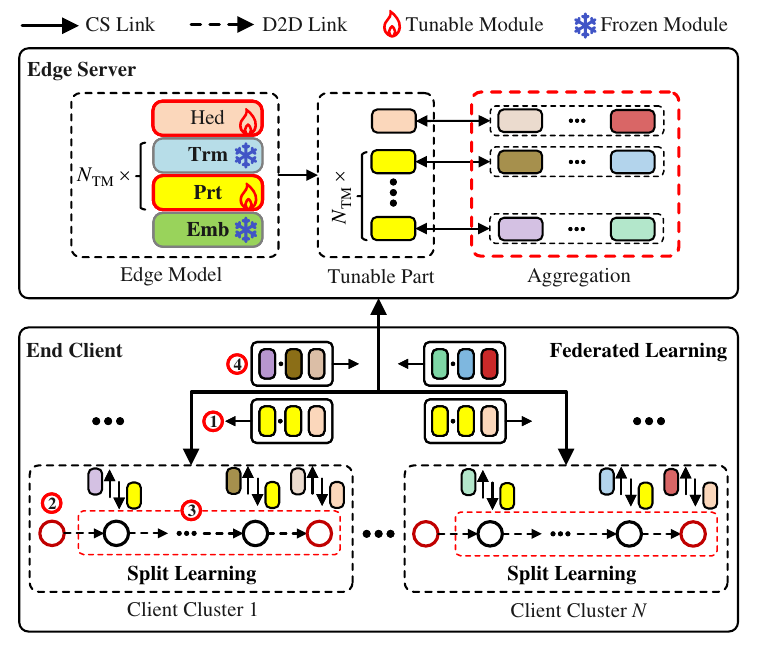}
\caption{The Framework of HFSL-based Model Fine-tuning.}
\label{fig_4}
\end{figure}

GaisNet guarantees model fine-tuning of GAI by FL-based inter-cluster parallel collaboration and SL-based intra-cluster serial collaboration. The proposed HFSL-based model fine-tuning framework is shown in Fig. \ref{fig_4}. Firstly, the parallel collaborative edge learning between the edge server and multiple client clusters is implemented in the FL framework to achieve privacy protection without data sharing and parameter-efficient model fine-tuning with low communication overhead, which could converge the terminal personalized knowledge produced by the fine-tuning client cluster to the edge model in the relevant domain. Then, serial collaboration is completed among clients within each fine-tuning client cluster in the mode of SL, where smashed data and gradients of learnable modules are transmitted between clients in the form of D2D communication. Ignoring the feedback of the small amount of gradient, the workflow of model fine-tuning in GaisNet includes processes composed of edge model segmentation and delivery, sensing data generation and embedding, computing and transmission of tunable modules, and uploading and aggregation of the terminal model.

\textit{1) Summary Workflow}

$\bullet$ \textbf{Segmentation and distribution of edge models.} The edge server splits the tunable part of the domain-specific edge model into prompt modules and one MLP head module. The above sub-modules are then delivered to different client clusters, where the members of each cluster undertake the complete tunable modules of an edge model. It is worth noting that when the clients’ computing resources are sufficient, consecutive prompt modules can be divided together and sent to the same client. If only one client cluster participates in the fine-tuning process, HFSL degenerates to SL.

$\bullet$ \textbf{Generation and embedding of training data.} The start point in each fine-tuning client cluster obtains the personalized labeled training data based on the sensing process, and then extracts the original features based on the local synchronous embedding layer.

$\bullet$ \textbf{Computing and transmission of tunable modules.} The output tokens of the start point that provides data are passed to the second client, and then the remaining clients in the fine-tuning client cluster complete the training of their responsible prompt modules utilizing serial D2D communication with the local fragmented computing resources. It consumes communication resources to transmit smashed data between the nodes in the cluster, including the forward tokens and the reverse gradients. It is worth noting that when the end clients’ computing resources are limited, the edge server can also act as a cooperative node, and part of the fine-tuning tasks can be executed at the edge server.

$\bullet$ \textbf{Uploading and aggregation of end model.} All fine-tuning client clusters upload the complete components of the edge model to the edge server, and then Fedavg-based parameter aggregation is performed among the same modules of different clusters, including prompt modules and MLP head modules with the same number. After aggregation, prompt modules and MLP head module in the edge model are updated.

The above process iterates until the model converges or a predefined number of rounds is reached.

\textit{2) Key Metrics}

Next, we summarize several key metrics of HFSL-based model fine-tuning in GaisNet \cite{ref21}.

$\bullet$ \textbf{Model performance} refers to the performance of task inference performed by the fine-tuned model, such as the accuracy of image recognition, which is the fundamental goal and measure metric of model fine-tuning and task inference of GaisNet.

$\bullet$ \textbf{Convergence latency} is the total latency required for model fine-tuning, including the time consumption of edge model delivery, sensing data generation, local model training (including parameter transmission based on D2D communication), and local model upload.

$\bullet$ \textbf{Computing cost} refers to the sum of computing power consumed by the participating clients for model training during the fine-tuning process.

$\bullet$ \textbf{Energy cost} includes the total energy consumption of the whole fine-tuning process. Similar to the convergence latency, it involves the energy consumption related to data acquisition, parameter transmission, and model training process.

$\bullet$ \textbf{Communication overhead} is the consumption of spectrum resources of the workflow involved in data transmission in the fine-tuning process, including the related resource consumption of client-server (CS) links and D2D links.

$\bullet$ \textbf{Memory footprint} is the sum of memory on each worker node used to temporarily store intermediate parameters and smashed data during the fine-tuning process.

\subsection{SL-based Task Inference}

\begin{figure}[t]
\centering
\includegraphics[width=3.0in]{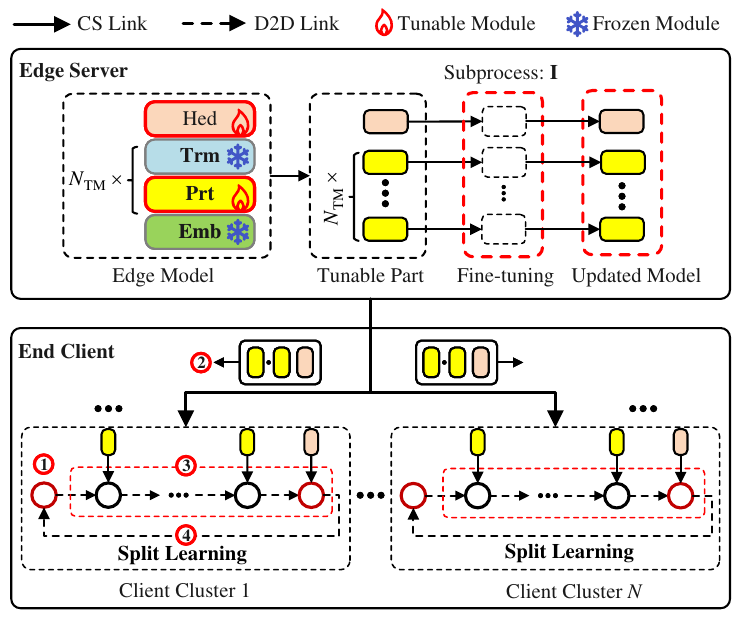}
\caption{The Framework of SL-based Task Inference.}
\label{fig_5}
\end{figure}

GaisNet can perform task inference via SL-based serial collaboration within the inference client cluster, and the proposed SL-based task inference framework is shown in Fig. \ref{fig_5}. A client with task inference service requirements will take itself as the start point and establish an inference client cluster containing multiple working clients. Then, the relevant domain edge server will send the updated modules after fine-tuning and aggregation to the working clients. Next, like the fine-tuning process, the clients in the inference client cluster perform a collaborative inference based on D2D communication. Finally, the end point feeds back the inference results output by the MLP head module to the start point that initiates the inference service.

\textit{1) Summary Workflow}

$\bullet$ \textbf{Generation and embedding of inference task.} The start point in each inference client cluster obtains the unlabeled sensing data as the inference task and then performs the original feature extraction of the sensing data based on the local synchronous embedding layer.

$\bullet$ \textbf{Segmentation and distribution of edge models.} The edge server splits the tunable part of the fine-tuned edge model into prompt modules and one MLP head module. Then the above sub-modules are sent to the inference client cluster, where the other members except the start point undertake the inference task of a complete edge model.

$\bullet$ \textbf{Computing and transmission of tunable modules.} The output tokens of the start point with inference service requirements are passed to the second client. Then, the remaining clients in the cluster complete the inference computing of their responsible modules by using the local fragmented computing resources connected via serial D2D communication. Notably, when the computing resources of the end clients are limited, the edge server can also act as a cooperative node, and part of the inference tasks can be performed at the edge server.

$\bullet$ \textbf{Output and feedback of inference results.} The end point in the inference client cluster outputs the inference result and then sends it back to the start point that initiates the inference service via the D2D communication link.

\textit{2) Key Metrics}

Next, we summarize several key metrics of SL-based task inference in GaisNet.

$\bullet$ \textbf{Model performance} is the fundamental goal and measure metric of model fine-tuning and task inference of GaisNet.

$\bullet$ \textbf{Inference latency} is the total delay during the whole inference process from the initiation of the inference task to the reception of the inference result.

$\bullet$ \textbf{Computing cost} refers to the sum of computing power consumed by the working clients involved in the inference process.

$\bullet$ \textbf{Energy cost} includes the total energy consumption of the whole inference process, including data acquisition, the parameters transmission, and the model inference process.

$\bullet$ \textbf{Communication overhead} is the consumption of spectrum resources of the workflow involved in data transmission in the inference process.

$\bullet$ \textbf{Memory footprint} is the amount of memory on each worker node used to temporarily store smashed data during inference, which is usually less than the memory required for fine-tuning.

\section{Major Issues for GaisNet}

\subsection{How to split model?}

Whether model fine-tuning or task inference, the difficulty of partitioning the tunable part in the edge GAI model is to determine the number of segmentation blocks and the location of segmentation points \cite{ref21}. Firstly, the number of blocks to be segmented needs to be adapted to the clients that can provide collaboration, and the block size of model segmentation, that is, how many tunable modules to be segmented into the same block, needs to be adapted in equal proportion to the resources of the corresponding clients that undertake the fine-tuning (inference) tasks of the block. Meanwhile, we need to consider the computing resources of the working clients, as well as the communication resources required for model delivery and uploading.

\subsection{How to cluster clients?}

Depending on the GAI service performed, we refer to the cluster formed by clients performing HFSL-based model fine-tuning as fine-tuning client cluster, and those performing task inference as inference client cluster. For both, it is necessary to consider the computing resource stock of the client, the linkability feature between the edge server and the client, and the communication resource demand, because the edge model needs to be sent to the terminal clients and where computing is performed \cite{ref13}. Meanwhile, the adjacent clients of the computing tasks in the cluster could communicate with the D2D links as they need to transfer the smashed data. In particular, for the fine-tuning client cluster, we need to consider the communication resources required for uploading the local model to the edge server, while for the inference client cluster, the D2D linkability between the end point and the start point needs to be satisfied because the inference results need to be fed back to the client initiating inference services. In addition to the above process, the start point of the fine-tuning client cluster needs to consider the availability and the quality of the generated data samples for training \cite{ref8}.

\subsection{Fine-tuning or Inference?}

Considering the limited resources of edge servers and end clients, simply and representatively, we assume that each round can only complete one GAI service, which can be a model fine-tuning service or a task inference service. GaisNet's model fine-tuning determines the performance upper bound for subsequent task inference, while inference results are the key measure metric of the quality of model fine-tuning. Taking the commodity production process as an example, model fine-tuning can be understood as upgrading equipment, while task reasoning is to create value by producing goods with equipment. Equipment upgrading does not create immediate revenue but needs to pay extra overhead. If the equipment is used to produce goods in subsequent rounds, the output can be increased, and the corresponding GaisNet can obtain better task inference performance. Therefore, \textit{model fine-tuning focuses on improving future benefits, while task inference determines immediate benefits}, and we need to realize the reasonable tradeoff to maximize the long-term cumulative benefits \cite{ref16,ref17,ref20}.

\subsection{Who does it serve?}

Generally, there are multiple edge server-initiated model fine-tuning services and client-initiated task inference services in GaisNet. Based on deciding fine-tuning or inference, due to limited resources, we need to consider which edge model to perform fine-tuning or which client to perform task inference for. For the former, we need to consider the adaptability of clients’ data to edge models of different domains and further consider the future service demand, i.e., which model will be more used in the future inference service, to select the edge model with the largest market demand and the largest performance gain generated by fine-tuning to perform fine-tuning. For the latter, we need to consider the relationship between the result of inference and the resource cost and select the inference service of the client with the largest profit to execute.

\section{Experimental Results and Discussions}

In this section, to prove the effectiveness of the proposed GaisNet framework and explore the relevant impact factors, we set the Transformer-based ViT-Base/16 model on the flower classification dataset as a case study \cite{ref3, ref4}. Our experiments are performed on the NVIDIA RTX4060 GPU platform with a learning rate of 0.001, the fine-tuned batch size is set to 10, and the training-validation ratio of the dataset is 4:1.

\subsection{Impact of Pre-training}

\begin{figure}[t]
\centering
\includegraphics[width=3.3in]{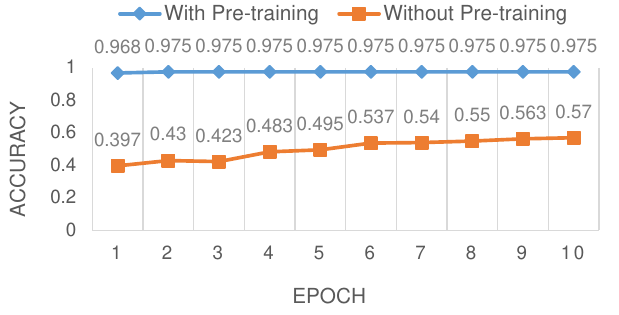}
\caption{The accuracy comparison between fine-tuning with pre-training and without.}
\label{fig_6}
\end{figure}

Pretraining can bring a strong prior foundation to edge models as well as to end clients. We compare the classification accuracy of the fine-tuning based on the pre-trained model with fine-tuning directly on the clients' local dataset ignoring the pre-trained model in the cloud. As shown in Fig. \ref{fig_6}, under the condition of using pre-trained FM, the accuracy of the first epoch can reach 96.80$\%$, which is significantly higher than 57.00$\%$, which is the convergence result without pre-training.

\subsection{Impact of Fine-tuning}
As shown in Fig. \ref{fig_6}, when we use the pre-trained model on the edge server, after 10 epochs of fine-tuning, the accuracy of the model is improved from 96.80$\%$ to 97.50$\%$, which realizes the adaptation to local data and performance enhancement. However, since the baseline is pre-trained on large-scale cloud data, the performance improvement that few-shot fine-tuning can achieve is limited.

\subsection{Impact of Parameter-efficient Fine-tuning}

\begin{figure}[t]
\centering
\includegraphics[width=3.5in]{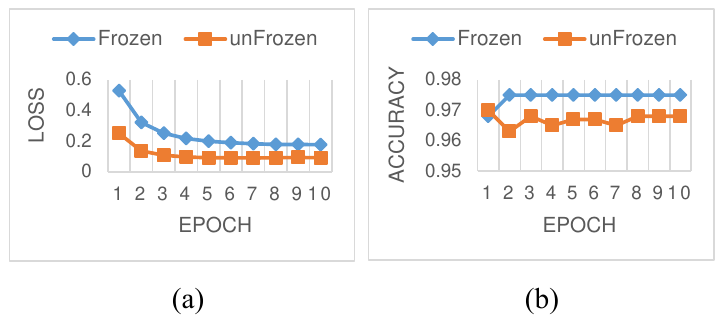}
\caption{The accuracy comparison between fine-tuning with frozen backbone and without.}
\label{fig_7}
\end{figure}

Freezing the backbone of the GAI model can significantly reduce the number of parameters for fine-tuning and reduce the computational resource overhead. As shown in Fig. \ref{fig_7}, the parameter-efficient fine-tuning method with the frozen backbone can converge to a higher accuracy after few-shot fine-tuning compared with the full parameter fine-tuning without backbone freezing. Meanwhile, the training time for each epoch in parameter-efficient fine-tuning is about 35s, while it takes 3 minutes and 30 seconds for full parameter fine-tuning.

\subsection{Impact of Non-IID Data}

\begin{table}[htbp]
\center
\footnotesize
\caption{The effect of the number of data classes on GaisNet.}
\label{table_3}
\renewcommand\arraystretch{2.0}
\begin{tabular}{|c|c|c|c|c|c|}
\hline
\textbf{Num Class} & \textbf{1} & \textbf{2} & \textbf{3} & \textbf{4} & \textbf{5} \\ \hline
\textbf{\makecell[c]{Accuracy\\(First/End)}} & \makecell[c]{0.200/\\0.200} & \makecell[c]{0.397/\\0.397} & \makecell[c]{0.575/\\0.595} & \makecell[c]{0.753/\\0.767} & \makecell[c]{0.933/\\0.967} \\ \hline
\end{tabular}
\end{table}

Table \ref{table_3} lists the results of fine-tuning with different numbers of data classes in the flower classification dataset, which can reflect the influence of Non-IID data of clients on the convergence of the GAI model in GaisNet. We can find that under the condition of the same amount of fine-tuning data, the performance of the GAI model decreases significantly with the increase of the degree of Non-IID.

\subsection{Impact of the Number of Client Cluster}

\begin{table}[t]
\center
\footnotesize
\caption{The effect of the number of client cluster on GaisNetI.}
\label{table_4}
\renewcommand\arraystretch{2.0}
\begin{tabular}{|c|c|c|c|c|c|c|}
\hline
\textbf{Num Cluster} & \textbf{1} & \textbf{2} & \textbf{3} & \textbf{4} & \textbf{5} & \textbf{6} \\ \hline
\textbf{\makecell[c]{Accuracy\\(First/End)}} & \makecell[c]{0.930/\\\textbf{0.950}} & \makecell[c]{0.940/\\\textbf{0.955}} & \makecell[c]{0.943/\\\textbf{0.960}} & \makecell[c]{0.950/\\\textbf{0.963}} & \makecell[c]{0.966/\\\textbf{0.974}} & \makecell[c]{0.968/\\\textbf{0.975}} \\ \hline
\end{tabular}
\end{table}

Table \ref{table_4} shows the impact of the number of client clusters involved in fine-tuning on the convergence accuracy of the edge model. Since more clusters will bring more personalized local data, the convergence accuracy of fine-tuning gradually improves as the number of clusters increases. However, due to the characteristics of few-shot fine-tuning of GAI, the increase in data size has limited improvement in accuracy.

\subsection{Impact of Integrated Fine-tuning and Inference}

\begin{figure}[t]
\centering
\includegraphics[width=2.8in]{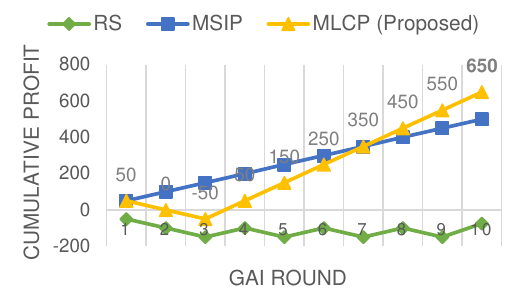}
\caption{The comparison of cumulative profit.}
\label{fig_8}
\end{figure}

\begin{table*}[t]
\center
\footnotesize
\caption{The service decisions and benefits.}
\label{table_5}
\renewcommand\arraystretch{2.0}
\begin{tabular}{|c|c|c|c|c|c|c|c|c|c|c|c|}
\hline
\textbf{IR/Task} & \textbf{1/A} & \textbf{2/A} & \textbf{3/B} & \textbf{4/C} & \textbf{5/C} & \textbf{6/C} & \textbf{7/C} & \textbf{8/C} & \textbf{9/C} & \textbf{10/C} & \textbf{ALL} \\ \hline
\textbf{RS} & a/-50 & b/-50 & a/-50 & C/50 & b/-50 & C/50 & a/-50 & C/50 & c/-50 & C/75 & \textbf{-75}\\ \hline
\textbf{MSIP} & \textbf{A/50} & \textbf{A/50} & \textbf{B/50} & C/50 & C/50 & C/50 & C/50 & C/50 & C/50 & C/50 & \textbf{500}\\ \hline
\textbf{\makecell[c]{MLCP\\(Proposed)}} & \textbf{A/50} & \textbf{c/-50} & \textbf{c/-50} & C/100 & C/100 & C/100 & C/100 & C/100 & C/100 & C/100 & \textbf{650}\\ \hline
\end{tabular}
\end{table*}

Adopting the concept of commodity production and equipment upgrading proposed in Section IV, we conduct a study on the impact of integrated fine-tuning and inference on GAI services in GaisNet. Suppose there are 3 devices (corresponding to the 3 edge models) producing 3 goods (corresponding to the 3 inference services that the edge models can provide). In each round, we can choose to upgrade one of the devices, denoted by a, b, and c, or choose one of the devices to generate goods, denoted by A, B, and C. We compare the proposed maximum long-term cumulative profit (MLCP) approach with integrated fine-tuning and inference with random selection (RS) and maximum the short-term immediate profit (MSIP) method, and the simulation results are shown in Table \ref{table_5}. Fig. \ref{fig_8} shows the change in cumulative profit with the GAI round. We can find that the proposed MLCP algorithm in the round 2 and the round 3 chooses to sacrifice the immediate profit to upgrade device c, to obtain more profit in the subsequent rounds, from which the maximum long-term cumulative profit can be obtained.

\section{Challenges and Future Directions}
\subsection{Privacy Concerns and Social Issues}
With the continuous application of GAI in various fields, it gradually raises concerns about privacy and social issues. First, the privacy leakage and data security issues involved in the process of pre-training, fine-tuning, and inference deserve attention. Second, we should attach importance to social concerns about issues such as prejudice, ethics, and intellectual property infringement. Therefore, it is necessary to study privacy protection strategies for GAI, such as differential privacy, enhance legislative supervision, and develop technologies that can enhance the traceability of GAI models, such as blockchain technology \cite{ref1}.

\subsection{Theoretical Bounds of GAI under Resource Constraints}
The performance of GAI is limited by the multi-domain physical resources and data with differentiated quantity and quality. Similar to Shannon's information theory, it is of great significance to study and describe the performance bounds of the GAI within limited resources and data for the development and deployment of GAI models.

\subsection{Incentive Mechanism Design}
Effective and fair incentive mechanisms are essential to encourage the wide participation of 6G end devices in the optimization of the GAI model. Comprehensive consideration of resource contribution and data differences ensures that the value created by each participant and the return obtained are matched \cite{ref7}. The mathematical tools of economics that can be utilized include game theory, auction theory, contract theory, and other theories \cite{ref11}.

\section{Conclusion}

In this paper, we propose GaisNet, a GAI-oriented collaborative cloud-edge-end intelligence framework, which dredges the bidirectional knowledge pipeline between GAI and EI and realizes efficient HFSL-based model fine-tuning and SL-based task reasoning on a unified architecture. Then we analyze the major issues in the running process of GaisNet and explore the effects of various influencing factors on GaisNet through simulation. Finally, we outline future challenges and directions in the interplay between GAI and EI.

\vfill


\begin{thebibliography}{1}
\bibliographystyle{IEEEtran}


\bibitem{ref1}
Y. Cao, S. Li, Y. Liu, Z. Yan, Y. Dai, P. Yu, L. Sun, ``A comprehensive survey of ai-generated content (aigc): A history of generative ai from gan to chatgpt,'' \textit{arXiv preprint arXiv}:2303.04226, 2023.

\bibitem{ref2}
A. Vaswani, N. Shazeer, N. Parmar, J. Uszkoreit, L. Jones, A. Gomez, L. Kaiser, I. Polosukhin, ``Attention is all you need,'' in \textit{Advances in neural information processing systems}, vol. 30, 2017.

\bibitem{ref18}
H. Du, Z. Li, D. Niyato, J. Kang, Z. Xiong, D. Kim and others, ``Enabling AI-generated content (AIGC) services in wireless edge networks,'' \textit{arXiv preprint arXiv:2301.03220}, 2023.

\bibitem{ref7}
W. Zhuang, C. Chen, L. Lyu, ``When foundation model meets federated learning: Motivations, challenges, and future directions,'' \textit{arXiv preprint arXiv:2306.15546}, 2023.


\bibitem{ref19}
G. Zhu, Z. Lyu, X. Jiao, P. Liu, M. Chen, J. Xu, S. Cui, P. Zhang, ``Pushing AI to wireless network edge: An overview on integrated sensing, communication, and computation towards 6G,'' in \textit{Science China Information Sciences}, vol. 66, no. 3, pp. 130301, 2023.


\bibitem{ref22}
X. Xia, F. Chen, Q. He, J. Grundy, M. Abdelrazek, and H. Jin, ``Online collaborative data caching in edge computing,'' \textit{IEEE Transactions on Parallel and Distributed Systems}, vol. 32, no. 2, pp. 281-294, 2020.

\bibitem{ref23}
X. Xia, F. Chen, Q. He, J. Grundy, M. Abdelrazek, and H Jin,  ``Cost-effective app data distribution in edge computing,'' \textit{IEEE Transactions on Parallel and Distributed Systems}, vol. 32, no. 1, pp. 31-44, 2020.



\bibitem{ref21}
S. Duan, D. Wang, J. Ren, F. Lyu, Y. Zhang, H. Wu, X. Shen, ``Distributed artificial intelligence empowered by end-edge-cloud computing: A survey,'' \textit{IEEE Communications Surveys \& Tutorials}, 2022.



\bibitem{ref11}
X. Huang, P. Li, H. Du, J. Kang, D. Niyato, D. Kim, Y. Wu, ``Federated Learning-Empowered AI-Generated Content in Wireless Networks,'' \textit{arXiv preprint arXiv:2307.07146}, 2023.


\bibitem{ref10}
Z. Zhang, Y. Yang, Y. Dai, Q. Wang, Y. Yu, L. Qu, Z. Xu, ``FedPETuning: When federated learning meets the parameter-efficient tuning methods of pre-trained language models,'' \textit{Association for Computational Linguistics (ACL)}, pp. 9963--9977, 2023.


\bibitem{ref12}
Y. Tian, Y. Wan, L. Lyu, D. Yao, H. Jin, L. Sun, ``FedBERT: When federated learning meets pre-training,'' in \textit{ACM Transactions on Intelligent Systems and Technology (TIST)}, vol. 13, no. 4, pp. 1--26, 2022.


\bibitem{ref15}
H. Zou, Q. Zhao, L. Bariah, M. Bennis, M. Debbah, ``Wireless multi-agent generative ai: From connected intelligence to collective intelligence,'' \textit{arXiv preprint arXiv:2307.02757}, 2023.

\bibitem{ref13}
Z. Lin, G. Qu, X. Chen, K. Huang, ``Split Learning in 6G Edge Networks,'' \textit{arXiv preprint arXiv:2306.12194}, 2023.



\bibitem{ref8}
A. Agarwal, M. Rezagholizadeh, P. Parthasarathi, ``Practical Takes on Federated Learning with Pretrained Language Models,'' \textit{Findings of the Association for Computational Linguistics: EACL 2023}, pp. 454--471, 2023.

\bibitem{ref4}
M. Jia, L. Tang, B. Chen, C. Cardie, S. Belongie, B. Hariharan, S. Lim, Ser-Nam, ``Visual prompt tuning,'' \textit{European Conference on Computer Vision}, pp. 709--727, 2022.

\bibitem{ref5}
J. He, C. Zhou, X. Ma, T. Berg-Kirkpatrick, G. Neubig, ``Towards a unified view of parameter-efficient transfer learning,'' \textit{arXiv preprint arXiv:2110.04366}, 2021.

\bibitem{ref6}
E. Hu, Y. Shen, P. Wallis, Z. Allen-Zhu, Y. Li, S. Wang, L. Wang, W. Chen, ``Lora: Low-rank adaptation of large language models,'' \textit{arXiv preprint arXiv:2106.09685}, 2021.


\bibitem{ref9}
J. Chen, W. Xu, S. Guo, J. Wang, J. Zhang, H. Wang, ``FedTune: A Deep Dive into Efficient Federated Fine-Tuning with Pre-trained Transformers,'' \textit{arXiv preprint arXiv:2211.08025}, 2022.

\bibitem{ref14}
Z. Cheng, X. Xia, M. Liwang, X. Fan, Y. Sun, X. Wang, L. Huang, ``CHEESE: distributed clustering-based hybrid federated Split learning over edge networks,'' \textit{IEEE Transactions on Parallel and Distributed Systems}, 2023.

\bibitem{ref3}
A. Dosovitskiy, L. Beyer, A. Kolesnikov, D. Weissenborn, X. Zhai, T. Unterthiner, M. Dehghani, M. Minderer, G. Heigold, S. Gelly and others, ``An image is worth 16x16 words: Transformers for image recognition at scale,'' \textit{arXiv preprint arXiv:2010.11929}, 2020.


\bibitem{ref16}
X. Li, S. Bi, H. Wang, ``Optimizing resource allocation for joint AI model training and task inference in edge intelligence systems,'' in \textit{IEEE Wireless Communications Letters}, vol. 10, no. 3, pp. 532--536, 2020.

\bibitem{ref17}
A. Eshratifar, M. Abrishami, M. Pedram, ``JointDNN: An efficient training and inference engine for intelligent mobile cloud computing services,'' in \textit{IEEE Transactions on Mobile Computing}, vol. 20, no. 2, pp. 565--576, 2019.

\bibitem{ref20}
N. Chen, Z. Cheng, X. Fan, B. Huang, X. Du, and G. Mohsen, ``Integrated Sensing, Communication, and Computing for Cost-effective Multimodal Federated Perception,'' \textit{arXiv preprint arXiv:2311.03815}, 2023.















\end{thebibliography}
\end{document}